\documentclass[11pt]{article}
\usepackage{mathtools}
\usepackage{amssymb}
\usepackage{amsmath}
\usepackage[T1]{fontenc}
\usepackage{slashed}
\usepackage{geometry}
\usepackage{graphicx}
\usepackage[bookmarksnumbered=true, hidelinks]{hyperref}
\usepackage{tikz}
\usepackage{color}
\usepackage{comment}
\usepackage{adjustbox}
\usepackage[ backend=bibtex, bibstyle=phys, citestyle=numeric-comp, sorting=none,  giveninits=true,  eprint=true,  doi=false,  url=false,  isbn=false,  articletitle=true,  biblabel=brackets]{biblatex}
\usepackage{microtype}

\newcommand{\g}{\gamma}
\newcommand{\spa}{\slashed{p}}
\newcommand{\sda}{\slashed{\partial}}
\newcommand{\sDa}{\slashed{D}}

\newcommand{\E}{\mathcal{E}}
\newcommand{\B}{\mathcal{B}}

\begin{document}
\begin{titlepage}
\begin{flushright}    
{\small $\, $}
\end{flushright}
\begin{center}
\vspace{2em}
\begin{adjustbox}{max width=\textwidth}
{\LARGE  \bf \scshape Planck mass gravitinos in  Einstein-Maxwell backgrounds }
\end{adjustbox}

\vspace{3em}

{\large \scshape Artur Krawczyk$^1$, Krzysztof A. Meissner$^1$,
Hermann Nicolai$^2$, \\[0.15em] and Bartłomiej Sikorski$^{1}$}
\vspace{1em}

\centerline{{\large $^1$} {Faculty of Physics, University of Warsaw}}
\centerline { {Pasteura 5, 02-093 Warsaw, Poland}}
\vskip 0.35cm
\centerline{{\large ${\,}^2$} {Max-Planck-Institut f\"ur Gravitationsphysik (Albert-Einstein-Institut)}}
\centerline { {Am M\"{u}hlenberg 1, 14476 Potsdam, Germany}}
\end{center}

\vspace{2em}
\begin{abstract}
\vspace{1em}
Charged massive spin-3/2 fields have long been regarded as problematic, because their coupling to electromagnetism generically leads to loss of hyperbolicity, acausal propagation and loss of unitarity. 
At the same time, fractionally charged supermassive gravitinos play a central role in a recent proposal of two of the present authors, where they are the only fermionic degrees of freedom beyond the Standard Model and provide novel dark matter candidates.
We address this apparent tension by revisiting the result of Deser and Waldron, who showed that the inclusion of gravity can compensate the electromagnetic source of the inconsistency. 
For a minimally coupled Rarita-Schwinger field in an Einstein-Maxwell background, consistency requires 
a lower bound on the mass in terms of the charge and the cosmological constant.
What is usually a severe obstruction to low-energy charged spin-3/2 phenomenology becomes an independent indication that such particles must lie close to the Planck scale, as argued by two of the present authors on other grounds.  We rederive the inequality from the solvability of the Rarita-Schwinger constraints and from the characteristic determinant, emphasizing how the Einstein-Maxwell stress tensor compensates the purely electromagnetic pathology. 
We then reformulate the system using a St\"uckelberg spinor, showing that the helicity-1/2 sector reproduces the same condition. 
This formulation is also essential for obtaining a renormalizable propagator.
We discuss this propagator and the corresponding ghost system.

\end{abstract}

\vfill

\thispagestyle{empty}
\end{titlepage}

\section{Introduction}

There is a long history of problems and paradoxes arising when one considers 
a charged massive spin-$\frac{3}{2}$ field coupled to electromagnetism outside of 
supergravity. Inconsistencies that arise with a minimal coupling include loss of unitarity \cite{Johnson:1960vt}, loss of hyperbolicity, and superluminal propagation \cite{Velo:1969bt}. 
Since the classic analysis of Velo-Zwanziger in \cite{Velo:1969bt}
the topic has been addressed in numerous publications over the 
past decades, sharpening the original analysis especially by including effects of gravity \cite{Madore1973, Madore:1975zz, Gibbons:1976fy, Deser:2000dzED, Deser:2001dt, Porrati:2009bs, Rahman:2011ik, Benakli:2026kue}. 
In particular, it was realized that the difficulties with purely electromagnetic 
couplings can be overcome when the coupling to gravity 
compensates the effects of electromagnetism. This is in particular the 
case for supergravity theories \cite{Deser:1976eh, Deser:1977uq}, 
but, subject to the inequality below,
remains true even without local supersymmetry, where again the 
inclusion of gravity in principle allows for a consistent theory.

More specifically, in \cite{Deser:2001dt} Deser and Waldron showed that the classical consistency of causal propagation of a Rarita-Schwinger field coupled minimally to electromagnetism and gravity requires an
inequality involving the mass and charge to be satisfied. To state
it we adopt units where the reduced Planck mass is the inverse of the gravitational coupling
$\kappa ^{-1}=M_{\mathrm{Pl}} =\sqrt{\frac{\hbar c}{8\pi G}}.$
For a given charge $q$,
\cite{Meissner:2020rzo, Meissner:2021nvx}
also introduced the notion of a `BPS mass' 
$M_{\rm BPS} \equiv M_{\rm BPS}(q)$, for which the electrostatic repulsion
between two particles of mass $M_{\rm BPS}$ and the same charge $q$
is balanced by their gravitational attraction, which in these units is
\begin{align}    \label{MBPS}
M_{\mathrm{BPS}}^2= \frac{q^2}{4\pi G}
=8\pi \alpha_q M_{\mathrm{Pl}}^2
\end{align} 
with $\epsilon_0=\mu_0=1$, $\alpha_q=\frac{q^2}{4\pi }$, and $c=\hbar =1$.
The relevant inequality then reads
\begin{align}\label{inequality}
   m^2>\frac{2 q^2}{3\kappa^2}-\frac{\Lambda }{3} 
   = \frac{8\pi\alpha_e}{3} \frac{q^2}{e^2}M_{\mathrm{Pl}}^2-\frac{\Lambda }{3} ,
\end{align}
where $m$ is the mass of the spin-$\frac32$ fermion, 
$\Lambda$ the cosmological constant, and 
$\alpha_e \sim (137)^{-1}$ is the usual fine structure constant.

Traditionally spin-$\frac 32$ fields are associated with local supersymmetry.
This ensures consistency \cite{Deser:1976eh, Deser:1977uq}
but raises questions with regard 
to the practical relevance of the above inequality  for physics beyond 
the Standard Model. In scenarios with low energy ($N=1$) supersymmetry, 
the gravitinos, being Majorana, cannot carry charges whence there 
is no problem with low values of $m$ even when the cosmological 
constant is tuned to zero (charged gravitinos 
require $N\geq 2$ supersymmetry). However, after decades of searches,
low energy supersymmetry is now almost excluded by collider experiments at LEP \cite{Ask:2003pg}, Tevatron \cite{PortellBueso:2011rnf} and LHC \cite{Sekmen:2025bxv}. Alternatively one may consider compactified
supergravity where after spontaneous breaking or Kaluza-Klein compactification
charged gravitinos can appear with very large masses, corresponding to
Planck scale breaking of supersymmetry. Again, (spontaneously broken) 
local supersymmetry ensures consistency, but now the vacuum is 
generically characterized by a huge negative cosmological constant, 
again in striking contradiction with observation. This dilemma was already 
emphasised in \cite{Deser:2001dt, Deser:1977uq}, apparently blocking any path towards 
implementing a realistic version of (spontaneously broken)
local supersymmetry in particle physics.

In this paper, we advocate a very different perspective, based on recent
work of two of the present authors \cite{Meissner:2018gtx, Meissner:2025qrx}.
This work originated from Gell-Mann's old observation that the 
spin-$\frac12$ fermions of maximal $N=8$ supergravity, 
after the removal of eight Goldstinos, can be matched with 
the three generations of quarks and leptons (including right-chiral neutrinos) 
of the Standard Model \cite{GellMann1985Renormalizability}.
This observation has gained renewed relevance with the non-observation of
new fermionic degrees of freedom at LHC and earlier experiments.
However, in Gell-Mann's scheme there remains a mismatch in the electric 
charges that cannot be fixed within $N=8$ supergravity. In order to
achieve a matching one must therefore go beyond supersymmetry. The
proposal is thus to replace supersymmetry with $E_{10}$ symmetry, 
and $N=8$ supergravity with a `pre-geometric' one-dimensional $E_{10}/K(E_{10})$ sigma model (or some more elaborate variant of this model),
in a scheme that necessarily transcends space-time based 
quantum field theory \cite{Damour:2002cu}.
The embedding of the $N=8$ supergravity 
R-symmetry group SU(8) into the vastly larger `maximal compact' subgroup
$K(E_{10})$  of $E_{10}$ allows to rectify the charge mismatches of quarks 
and leptons \cite{Meissner:2014SMfermions, Kleinschmidt:2015sfa}.
An important distinction between the $E_{10}$ model and 
$N=8$ supergravity is that the $E_{10}$ model does not allow 
for static space-time geometries such as the AdS vacua common 
in gauged supergravities, but necessarily requires time-dependent 
cosmologies \cite{Kleinschmidt:2005gz}. Thus, it is clearly
incompatible with space-time supersymmetry.

A central feature of this scheme is that the only other fermions beyond the
48 spin-$\frac12$ fermions of the Standard Model are eight massive gravitinos,
with fractional charges. Taking into account the gravitino charge shifts that
accompany the charge shifts of the spin-$\frac12$ fermions, 
these charges acquire values which protect the gravitinos from 
decaying into Standard Model fermions. For this reason,
they have been proposed as novel Dark Matter candidates \cite{Meissner:2018cayDM}. 
Consideration of bounds on the electric charges of Dark Matter particles \cite{McDermott:2010pa, Dolgov:2013una, DelNobile:2015bqo}  leads to the conclusion that these particles must be superheavy. Hence, their
abundance in the universe would have to be extremely low.

In this paper, we supply an extra and independent argument for large mass 
values, on the basis of purely theoretical consistency arguments encountered
in the works mentioned above. Namely, in the absence of supersymmetry 
the above inequality likewise imposes strict limits
on the allowed mass range of gravitinos. When the effective $\Lambda$ is put
equal to zero, (\ref{inequality}) becomes
\begin{align}\label{inequality0}
   m^2>\frac{2 q^2}{3\kappa^2}
   = \frac{8\pi\alpha_e}{3} \frac{q^2}{e^2}M_{\mathrm{Pl}}^2   \; .
\end{align}

Note that even the positive $\Lambda$ needed for inflation would not 
affect this inequality in any appreciable way because the relevant energy scale
is now known to be much below the Planck scale \cite{Guth:2013sya, Planck:2018jri}. 
In that case, and without supersymmetry, we see that it is this inequality 
that forces the gravitino mass to lie in a small mass range below the 
Planck mass. 

Here we want to point out that, up to a numerical factor of 3,
the above bound coincides with a bound derived on the basis of very 
different considerations, namely the requirement that, for sufficiently
heavy gravitinos, the gravitational attraction should overcome the electrostatic 
repulsion between two like-charged gravitinos thus enabling gravitational 
collapse of gravitino lumps in the very early universe. Whenever these 
lumps are large enough to overcome Hawking radiation, that is,
whenever the radiation temperature exceeds the Hawking temperature,
the collapsed lumps can grow rapidly and thus explain the formation of primordial black holes in the early universe  \cite{Meissner:2020rzo, Meissner:2021nvx}. The relevant inequality reads
\begin{align}\label{BPS_bound}
m^2 \,>\, M_{\mathrm{BPS}}^2  \; 
\end{align}
with $M_{\mathrm{BPS}}$ as in \eqref{MBPS}.  Remarkably, for the supergravity Pauli coupling considered in \cite{Deser:2001dt}, the causality bound coincides exactly with this BPS bound.

The bound derived in \cite{Deser:2001dt} is thus consistent with the independent 
analysis of primordial black hole formation in \cite{Meissner:2020rzo} and the 
conjectured role of gravitinos as dark matter candidates \cite{Meissner:2018cayDM}.
For gravitinos of charge $q=\frac{2}{3}e$, as in \cite{Meissner:2025qrx}, the inequality \eqref{inequality0} becomes
\begin{align}
    m\gtrsim 0.16 \ M_{\mathrm{Pl}} \simeq 4\cdot 10^{17}\ \mathrm{GeV}.
\end{align}

The notion of consistency relevant for massive spin-$\frac{3}{2}$ fields has two complementary aspects. The first is classical: the constrained Rarita-Schwinger system must possess algebraic constraints that can be solved on the chosen background, and its characteristic hypersurfaces must lie inside, or on, the metric light cone. 
This is the content of the bound reviewed below. The second aspect is perturbative: the massive spin-$\frac{3}{2}$ field should be described in variables for which the propagator has acceptable high-momentum behavior. 
In the ordinary massive Rarita-Schwinger formulation the propagator contains positive powers of momentum divided by powers of the mass, as in a unitary gauge of the massive vector field. 
Introducing a St\"uckelberg spinor restores a local fermionic gauge symmetry of the massless theory, isolates the helicity-$\frac{1}{2}$ sector responsible for the characteristic obstruction, and permits a Feynman gauge in which the action has a Dirac kinetic term for a vector-spinor. 
The price is the corresponding commuting spinor ghost system that we study using the BRST framework. Thus, the St\"uckelberg formulation provides both another derivation of the classical causality obstruction and a propagator suitable for standard perturbative power counting.
The main new contribution of the present work is the reinterpretation in the superheavy gravitino setting, together with a St\"uckelberg/BRST formulation that exposes the causality obstruction and yields a propagator suitable for perturbative power counting.

 Finally, it must be emphasized that these considerations apply
only to elementary and stable charged spin-$\frac32$ particles in the 
Rarita-Schwinger setting with electromagnetic and gravitational backgrounds. 
Indeed,  the spin-$\frac32$ $\Omega^-$ resonance, being charged and with 
a mass far below the Planck scale, would seem to provide an obvious 
counterexample to (\ref{inequality0}). However, that particle is composite, 
subject to complicated strong interaction dynamics, and highly
unstable; see also \cite{Benakli:2026kue} for a discussion of this point.

The paper is organized as follows. In Section \ref{sec:causal}, we derive the constraint and characteristic conditions for a minimally coupled charged Rarita-Schwinger field on an Einstein-Maxwell background and recover the Planckian bound of Deser-Waldron \cite{Deser:2001dt}. 
In Section \ref{sec:stuck}, we reformulate the same system using a St\"uckelberg spinor and show that the helicity-$\frac{1}{2}$ sector reproduces the same characteristic obstruction and leads to the same consistency condition as the analysis of constraints.
Section \ref{sec:propagator} discusses the propagator of the massive Rarita-Schwinger field, the gauge fixing that brings the St\"uckelberg theory to a Dirac-like form with a renormalizable propagator, and associated ghost action.

\section{Causality of Planck mass gravitinos}\label{sec:causal}
We consider a charged massive Rarita–Schwinger field\footnote{Our signature is $(+,-,-,-)$. For $\gamma$-matrices we use
$ \{\gamma^\mu,\gamma^\nu\}=2g^{\mu\nu},$ $
\gamma^{\mu_1\cdots\mu_n}\equiv \gamma^{[\mu_1}\cdots\gamma^{\mu_n]}, $
with antisymmetrization of unit weight and
$
\gamma^5=i\gamma^0\gamma^1\gamma^2\gamma^3,$ $
\epsilon^{0123}=+1/\sqrt{-g}.$} minimally coupled to background fields in an Einstein-Maxwell background
\begin{align}\label{eq:Lagrangian_RS}
 \frac{1}{\sqrt{-g}}\mathcal{L}_{}=-i\bar{\psi}_\mu \g^{\mu\nu\rho}D_\nu \psi_\rho+m\bar{\psi}_\mu\gamma^{\mu\nu}\psi_\nu.
\end{align}
We take $\psi_\mu$ to have a charge $q$, and the covariant derivative acts on it as 
\begin{align}
    D_\mu\psi_\nu = \partial_\mu \psi_\nu-\Gamma^{\rho}{}_{\mu\nu}\psi_\rho+\frac{1}{4}\omega_{\mu ab}\g^{ab}\psi_\nu+iqA_\mu\psi_\nu.
\end{align}
The field strength reads
\begin{align}
    [D_\mu,D_\nu]\psi_\rho =-R_{\mu\nu\rho}\ ^\sigma \psi_\sigma+\frac{1}{4}R_{\mu\nu ab}\gamma^{ab}\psi_{\rho}+iqF_{\mu\nu}\psi_\rho,
\end{align}
and it will always appear in the fully antisymmetrized form
\begin{align}
    D_{[\mu}D_\nu\psi_{\rho]}=\frac{1}{4}R_{[\mu\nu |ab|}\gamma^{ab}\psi_{\rho]}+iq F_{[\mu\nu}\psi_{\rho]}.
\end{align}
We now show that in an Einstein-Maxwell background the gravitational contribution modifies the constraint and characteristic equations so that the Velo--Zwanziger obstruction is absent for masses above the Planckian bound. Our analysis closely follows the one performed by Deser and Waldron \cite{Deser:2001dt} and rederives their results.

\subsection{Solvability of constraints}

The Rarita-Schwinger equation is a constrained system and, depending on the background configuration, constraints may not be solvable.
Equations of motion following from \eqref{eq:Lagrangian_RS} are
\begin{align}\label{eom}
    E^\mu\equiv -i\gamma^{\mu\nu\rho}D_\nu \psi_\rho +m\gamma^{\mu\nu}\psi_\nu =0,
\end{align}
which upon a contraction with $\gamma^\mu $ implies
\begin{align}
i\gamma^{\nu\rho}D_\nu\psi_\rho = \frac{3}{2}m\,\gamma^\rho\psi_\rho .
\end{align}
Upon choice of a timelike variable $\mu=0$, we see that the spatial components $\psi_i$ are the dynamical variables, while $\psi_0$ is non-dynamical (there is no $\dot{\psi}_0$ in \eqref{eom}) and is determined from constraints.
To solve for $\psi_0$, we take the divergence of \eqref{eom}
\begin{align}
    2iD_\mu E^\mu=\g_\mu G^{\mu\nu}\psi_\nu +iqF_{\mu\nu}\gamma^{\mu\nu\rho}\psi_\rho +2im\g^{\mu\nu}D_\mu \psi_\nu=\g_\mu G^{\mu\nu}\psi_\nu +2q\gamma_\mu \gamma^5 \tilde{F}^{\mu\nu}\psi_\nu+3m^2\gamma ^\mu\psi_\mu,
\end{align}
where we used the identity 
\begin{align}
    \gamma^{\mu\nu\rho}=i\epsilon^{\mu\nu\rho\delta}\g_\delta \gamma^5.
\end{align}
The coefficient that multiplies the non-dynamical component $\psi_0$ is therefore a spinor matrix
\begin{align}\label{eq:Matrix_psi0}
 \mathcal{M}=  \g_\mu G^{\mu 0}+2q\gamma_\mu \gamma^5 \tilde{F}^{\mu 0}+3m^2\gamma ^0,
\end{align}
whose determinant can be computed using the identity
\begin{align}
    \det \left[\gamma^\mu (a_\mu+\gamma^5b_\mu)  \right] = (a+b)^2(a-b)^2,
\end{align}
so it takes the value
\begin{align}\label{determinant_with_gravity}
    \det \mathcal{M}\propto \left(\delta_0^\mu+
    \frac{1}{3m^2}G^{\mu 0}+ \frac{2q}{3m^2}\tilde{F}^{\mu 0}
    \right)^2\left(\delta_0^\mu+
    \frac{1}{3m^2}G^{\mu 0}-\frac{2q}{3m^2}\tilde{F}^{\mu 0}
    \right)^2.
\end{align}
First, observe that we can assume the cosmological constant can be put to vanish by shifting $ m^2\mapsto m^2+\Lambda/3$, 
so it can be put to $0$ if $\Lambda>-3m^2$ and restored later\footnote{We can see that for $\Lambda<-3m^2$ there is a superluminal propagation even for a vanishing electromagnetic field. This case is tachyonic and nonunitary \cite{Deser:2001AdS}, so we don't consider it here.}.
We write the Einstein tensor for the Einstein-Maxwell system using $\vec{E}$ and $\vec{B}$ fields in a given frame
\begin{align}
    G^{00}&=\kappa^2 T^{00}=\frac{\kappa^2}{2}(\vec{E}^2+\vec{B}^2), \\
    G^{0i}&= \kappa^2 T^{0i}=\kappa^2(\vec{E}\times\vec{B})^i,\\
   G^{ij} &= \kappa^2 T^{ij}=\kappa^2\left(B^iB^j +E^iE^j-\frac{1}{2}(\vec{E}^2+\vec{B}^2)\delta^{ij} \right) .
\end{align}
The determinant of \eqref{determinant_with_gravity} vanishes when
\begin{align}
  1+\frac{\kappa^2(\vec{E}^2+\vec{B}^2)}{3 m^2}+\frac{\kappa^4(\vec{E}^2+\vec{B}^2)^2}{36 m^4}- \frac{\kappa^4}{9 m^4}\sin^2\theta\vec{E}^2\vec{B}^2 -\frac{4q^2}{9m^4}
  \vec{B}^2
  &=0,
\end{align}
where $\theta$ is the angle between $\vec{E}$ and $\vec{B}$.
We can introduce new variables $\E=\frac{\kappa^2}{6 m^2}\vec{E}^2$, $\B=\frac{\kappa^2}{6 m^2}\vec{B}^2$ to write the above as
\begin{align}
    (1+\E +\B)^2-4\sin^2\theta \E\B -\frac{8 q^2}{3\kappa^2m^2}\B= 0.
\end{align}
 Then the l.h.s. of the above equation is bounded below by
\begin{align}
 (1+ \E +\B)^2-4\E\B -\frac{8 q^2}{3\kappa^2m^2}\B =
 (1 + \E -\B)^2 + 4\B\left( 1 - \frac{2 q^2}{3\kappa^2 m^2}\right)
\end{align}
whose positivity is ensured by the inequality 
\begin{align}
 1 - \frac{2 q^2}{3\kappa^2 m^2} \,> \, 0
 \end{align}
independently of the values of $\E$ and $\B$. By putting $\Lambda$ back
we arrive at the more general inequality \eqref{inequality} that ensures positivity of the determinant \eqref{determinant_with_gravity}.

In the absence of gravitational coupling, i.e. for the Maxwell background in flat spacetime, the matrix in \eqref{eq:Matrix_psi0} is
\begin{align}
      \mathcal{M}=3\gamma^0\left(\frac{2q}{3}\gamma_\mu \gamma^5 \tilde{F}^{\mu 0} +m^2\right),
\end{align}
which is degenerate when $\vec{B}^2=\left(\frac{3m^2}{2q}\right)^2$, 
as first shown by Velo and Zwanziger \cite{Velo:1969bt}.

\subsection{Analysis of characteristics}
We next analyze the principal symbol of the field equation. This addresses a question different from algebraic solvability of $\psi_0$. A characteristic hypersurface is defined by the vanishing of the determinant of the principal symbol. If the corresponding normal $n_\mu$ can be timelike with respect to the background metric, then wavefronts describing propagation lie outside the metric light cones. Hyperbolicity is a stronger condition: for fixed spatial covector $\vec n$, the characteristic equation should admit only real values of $n_0$.

Using the notation of \cite{Madore1973, Deser:2000dzED, Deser:2001dt} we can write the discontinuity of field equations
\begin{align}
    [-i\partial_\mu \psi_\nu]=n_\mu \Psi_\nu, 
\end{align}
which is given by a spinor-vector solution $\Psi_\nu$.
We can write discontinuities of equations of motion
\begin{align}
    [E^\mu ]&=\gamma^{\mu\nu\rho}n_\nu\Psi_\rho=0,\\
   \frac{1}{2} [\gamma_\mu E^\mu ]&=\gamma^{\nu\rho}n_\nu\Psi_\rho=\slashed{n}\gamma\cdot \Psi -n\cdot\Psi=0,\\
   \left[E^\mu-\frac{1}{2}\g^\mu\gamma_\nu E^\nu \right]&=\slashed{n}\Psi^\mu-n^\mu \gamma\cdot \Psi=0,
\end{align}
from which follows\begin{align}
    n^2\Psi_\mu = n_\mu n\cdot \Psi.
\end{align}
We look at the discontinuity of the constraint $D_\mu E^\mu$
\begin{align}
   2n^\alpha[\partial_\alpha D_\nu E^\nu]&=n^\alpha\left(\g_\mu G^{\mu\nu}n_\alpha +2q\gamma_\mu \gamma^5 \tilde{F}^{\mu\nu}n_\alpha+3m^2\gamma ^\nu n_\alpha\right)\Psi_\nu\\
   &=\left(\g_\mu G^{\mu\nu}n_\nu +2q\gamma_\mu \gamma^5 \tilde{F}^{\mu\nu}n_\nu+3m^2\gamma ^\nu n_\nu\right)n\cdot\Psi=0 \label{discont_eq_matrix}.
\end{align}
The existence of the nontrivial solution $\Psi_\mu$ amounts to the vanishing of the following determinant
\begin{align}
  \det & \left( \slashed{n} +\frac{1}{3m^2}G^{\mu\nu}\gamma_\mu n_\nu+\frac{2q}{3m^2}\tilde{F}^{\mu\nu}\g_\mu\gamma^5n_\nu\right) 
  \notag
  \\=&\left(n^\mu +\frac{1}{3m^2}G^{\mu\nu}n_\nu-\frac{2q}{3m^2}\tilde{F}^{\mu\nu}n_\nu\right)^2\left(n^\mu+\frac{1}{3m^2}G^{\mu\nu}n_\nu +\frac{2q}{3m^2}\tilde{F}^{\mu\nu}n_\nu\right)^2=0.
\end{align}
Hence, we check if there are timelike solutions to
\begin{align}\label{determinant_gravity_EM}
0&=\left( n^\mu +\frac{1}{3m^2}G^{\mu\nu} n_\nu\pm \frac{2q}{3m^2}\tilde{F}^{\mu\nu} n_\nu\right)^2\notag \\&=n^\mu n_\mu +\frac{1}{9m^4}G^{\mu\alpha}G_{\alpha}{}^{\nu}n_\mu n_\nu+\frac{2}{3m^2}G^{\mu\nu}n_\mu n_\nu-\frac{4q^2}{9m^4}\tilde{F}^{\mu\alpha}\tilde{F}_{\alpha}{}^{\nu}n_\mu n_\nu\pm \frac{4q}{9m^4}G^{\mu\alpha}\tilde{F}_{\alpha}{}^{\nu}n_\mu n_\nu.
\end{align}
If the equation \eqref{determinant_gravity_EM} has a timelike solution $n_\mu=(n_0,\vec{n})$, then the system admits a characteristic hypersurface whose normal is timelike, corresponding to propagation outside the metric light cone.

In an Einstein-Maxwell background
\begin{align}
    G^{\mu\nu}=\kappa^2 T^{\mu\nu}+\Lambda g^{\mu\nu},
\end{align}
where $T^{\mu\nu}$ is the Maxwell stress-energy tensor. The term $G^{\mu\alpha}\tilde{F}_{\alpha}{}^{\nu}n_\mu n_\nu$ vanishes due to antisymmetry of $G\tilde{F}$ in this background.
Again, the cosmological constant in \eqref{determinant_gravity_EM} can be shifted to $0$ and restored later. Using
\begin{align}
\tilde{F}^{\mu\alpha}\tilde{F}_{\alpha}{}^{\nu}
=  T^{\mu\nu}+\frac{1}{2}g^{\mu\nu}(\vec{B}^2-\vec{E}^2),
\end{align}
we can write the expression in terms of the stress tensor and relativistic invariants
\begin{align}
     0&=n^\mu n_\mu +\frac{1}{9m^4}G^{\mu\alpha}G_{\alpha}{}^{\nu}n_\mu n_\nu+\frac{2}{3m^2}G^{\mu\nu}n_\mu n_\nu-\frac{4q^2}{9m^4}\tilde{F}^{\mu\alpha}\tilde{F}_{\alpha}{}^{\nu}n_\mu n_\nu\\
     &=n_\mu \left(g^{\mu\nu} +\frac{\kappa^4}{9m^4}T^{\mu\alpha}T_{\alpha}{}^{\nu}+\frac{2\kappa^2}{3m^2}T^{\mu\nu}\left(1- \frac{2 q^2}{3\kappa^2 m^2} \right)+\frac{2q^2}{9m^4}g^{\mu\nu}(\vec{E}^2-\vec{B}^2) \right)n_\nu\\
      &=n_\mu \left(g^{\mu\nu}\left(1+\frac{2q^2}{9m^4}(\vec{E}^2-\vec{B}^2) + \frac{\kappa^4}{9 m^4}\left((\vec{E}\cdot \vec{B})^2 +\frac{1}{4}(\vec{E}^2-\vec{B}^2)^2\right)\right) +\frac{2\kappa^2}{3m^2}T^{\mu\nu}\left(1- \frac{2 q^2}{3\kappa^2 m^2} \right) \right)n_\nu .
\end{align}
We see that if $m^2>2 q^2/3\kappa^2$, then there is no timelike solution as both $ n_\mu g^{\mu\nu}n_\nu$ and $ T^{\mu\nu}n_\mu n_\nu$ are positive, and their coefficients are positive for all values of $\vec{E}$ and $\vec{B}$. Otherwise, the coefficient of the Maxwell stress-tensor contribution changes sign, and explicit electromagnetic backgrounds can be chosen for which the characteristic equation admits timelike normals. 
For nonzero $\Lambda$, the bound is $m^2>2 q^2/3\kappa^2 - \Lambda/3$, which is the same condition as the one derived from the solvability of constraints.

We now show that without the gravitational coupling there is always a superluminal propagation of constraints, which is given by a spacelike direction in the characteristic cone. This follows from the existence of a timelike $(n_0,\vec{n})$ normal to the characteristic surface.  
The equation \eqref{discont_eq_matrix} contains the matrix which we can write in the Weyl basis
\begin{align}
    &\mathcal{M}=\slashed{n} +\frac{2q}{3m^2}\tilde{F}^{\mu\nu}\g_\mu\gamma^5n_\nu=\\& \begin{pmatrix}
        0 &( n_0 -\frac{2q}{3m^2}\vec{n}\cdot\vec{B}) -\vec{\sigma}\cdot(\vec{n} -\frac{2q}{3m^2}n_0\vec{B}+\frac{2q}{3m^2}\vec{n}\times \vec{E})\\ ( n_0 +\frac{2q}{3m^2}\vec{n}\cdot\vec{B}) +\vec{\sigma}\cdot(\vec{n} +\frac{2q}{3m^2}n_0\vec{B}+\frac{2q}{3m^2}\vec{n}\times \vec{E})& 0
    \end{pmatrix}.\notag
\end{align}
For $\vec{n}$ in the direction of $\vec{E}$, the determinant becomes
\begin{align}
     \det \mathcal{M}_{\vec{n}\| \vec{E}} 
    &=\left( n_0^2\left(1-\frac{4q^2}{9m^4}\vec{B}^2\right)-\vec{n}^2\left(1-\cos^2\theta\frac{4q^2}{9m^4}\vec{B}^2 \right)\right)^2 . 
\end{align}
It vanishes for timelike vectors $(n_0, \vec{n})$ provided
\begin{align}\label{acausality_result_general_B}
\frac{n_0}{|\vec{n}|} = \sqrt{\frac{1-\cos^2\theta\frac{4q^2}{9m^4}\vec{B}^2}{
1-\frac{4q^2}{9m^4}\vec{B}^2
}},
\end{align}
and one can see that the equation can cease to be hyperbolic 
(can only be solved with imaginary $n_0$) when
\begin{align}
\vec{B}^2 > \left(\frac{3m^2}{2q}\right)^2.     
\end{align}
Furthermore, the above equation shows that for general values of $\theta$ we have $n_0 > |\vec{n}|$, hence the 4-vector $n^\mu$ orthogonal to the characteristic surface is timelike 
implying superluminal propagation.

\section{St\"uckelberg formulation}\label{sec:stuck}
We can now reformulate the above analysis using an auxiliary St\"uckelberg field. This field isolates the sector that exhibits superluminal propagation. Additionally, it is required to obtain the propagator in a form that decays with momentum suitably for loop calculations.

First, we consider the   gravitino-St\"uckelberg system in flat spacetime and in the absence of electromagnetic interactions. We do this first to study kinetic terms. This also serves as a foundation for quantum calculations around Minkowski vacuum, where all interaction terms are assumed to be small to be treated perturbatively.

The Lagrangian of the massless Rarita-Schwinger field in flat spacetime
\begin{align}
\mathcal{L}_{m=0}=-i\Bar{\psi}_\mu \g^{\mu\nu\rho}\partial_\nu \psi_\rho
\end{align}
possesses a symmetry\footnote{We consider the Dirac field, so there are two independent symmetry parameters $\epsilon,\bar{\epsilon}$.
 }
\begin{align}\label{massless_transformation}
      \delta' \psi_\mu=\partial_\mu\epsilon', \qquad  \delta' \bar{\psi}_\mu=\bar{\epsilon}'\overleftarrow{\partial}_\mu
\end{align}
with local spin-$\frac{1}{2}$ fermions $\epsilon',\bar{\epsilon}'$. The quantization of this system was first described in \cite{Das:1976ct}.
When $m\neq 0$, the Lagrangian is not invariant under the transformation \eqref{massless_transformation}.

To restore symmetry we add a St\"uckelberg field $\chi$ that transforms as
\begin{align}
    \delta \chi =m\epsilon, \qquad  \delta \bar{\chi} =m\bar{\epsilon}.
\end{align}
For $\chi$ to have a kinetic term and avoid kinetic mixing,
the transformation rules for $\psi_\mu$ are modified to
\begin{align}
    \delta\psi_\mu = \left( \partial_\mu-i\frac{m}{2}\gamma_\mu\right)\epsilon, \qquad   \delta\bar{\psi}_\mu =\bar{\epsilon} \left( \overleftarrow{\partial}_\mu+i\frac{m}{2}\gamma_\mu\right).
\end{align}
and the invariant Lagrangian is 
\begin{align}\label{eq:inv_lagr}
    \mathcal{L} &= -i\Bar{\psi}_\mu \g^{\mu\nu\rho}\partial_\nu \psi_\rho+m\bar{\psi}_\mu\gamma^{\mu\nu}\psi_\nu +\frac{3}{2}\bar{\chi}(i\sda +2m)\chi+\frac{3}{2}im(\bar{\psi}\cdot \g\chi-\bar{\chi}\g\cdot \psi).
\end{align}
This Lagrangian can be derived using the invariant combination 
\begin{align}\label{lagrangian_Stuckelberg_free}
\psi_\mu'=\psi_\mu-\frac{1}{m}\left(\partial_\mu -i\frac{m}{2}\gamma_\mu\right)\chi    ,\quad \bar{\psi}_\mu'=\bar{\psi}_\mu-\frac{1}{m}\bar{\chi}\left(\overleftarrow{\partial}_\mu +i\frac{m}{2}\gamma_\mu\right)  .
\end{align}

\subsection{Causality of the St\"uckelberg system in Einstein-Maxwell background}
The system \eqref{lagrangian_Stuckelberg_free} has the gauge symmetry which replaces usual constraints so the gamma-trace part of $\psi_\mu$ is compensated by $\chi$. It is therefore natural to assume that the superluminal propagation of constraints will be reflected in equations for $\chi$. It was observed by Rahman \cite{Rahman:2011ik} that this helicity-$\frac{1}{2}$ mode probes the pathologies of minimal interactions found by Velo and Zwanziger.

The covariant derivative couples gravity and $A_\mu$ to the St\"uckelberg field by
\begin{align}\label{covariant_derivatives_CST}
        D_\mu\chi &= \partial_\mu \chi+\frac{1}{4}\omega_{\mu ab}\g^{ab}\chi+iqA_\mu \chi.
\end{align}
The Lagrangian is invariant under
\begin{align}
    \delta\psi_\mu = \left(D_\mu-i\frac{m}{2}\gamma_\mu\right)\epsilon, \qquad   \delta\bar{\psi}_\mu =\bar{\epsilon} \left( \overleftarrow{D}_\mu+i\frac{m}{2}\gamma_\mu\right), \qquad  \delta \chi =m\epsilon, \qquad  \delta \bar{\chi} =m\bar{\epsilon}.
\end{align}
so the invariant combination is
\begin{align}
\psi_\mu'=\psi_\mu-\frac{1}{m}\left(D_\mu -i\frac{m}{2}\gamma_\mu\right)\chi    ,\quad \bar{\psi}_\mu'=\bar{\psi}_\mu-\frac{1}{m}\bar{\chi}\left(\overleftarrow{D}_\mu +i\frac{m}{2}\gamma_\mu\right)  .
\end{align}
The Lagrangian now includes covariant derivatives and coupling to field strengths 
\begin{align}
   \frac{1}{\sqrt{-g}} \mathcal{L} &= -i\Bar{\psi}_\mu \g^{\mu\nu\rho}D_\nu \psi_\rho+m\bar{\psi}_\mu\gamma^{\mu\nu}\psi_\nu +\frac{3}{2}\bar{\chi}(i\sDa +2m)\chi+\frac{3}{2}im(\bar{\psi}\cdot \g\chi-\bar{\chi}\g\cdot \psi)+\mathcal{L}_{D^2}+\mathcal{L}_{D^3},
\end{align}
where terms with two and three commuted covariant derivatives are
\begin{align}
  \notag\mathcal{L}_{D^2}&=\frac{1}{m}\left(i\Bar{\psi}_\mu \g^{\mu\nu\rho}D_\nu D_\rho\chi-i\Bar{\chi}\g^{\mu\nu\rho}D_\mu D_\nu \psi_\rho+\frac{1}{2}\bar{\chi}\gamma_\mu\gamma^{\mu\nu\rho}D_\nu D_\rho\chi+\frac{1}{2}\bar{\chi}\gamma^{\mu\nu\rho}D_\mu D_\nu\g_\rho\chi-\bar{\chi}\gamma^{\mu\nu} D_\mu D_\nu\chi \right) \\
  \label{LD2}&=\frac{1}{2m}\left( qF_{\mu\nu}-\frac{i}{4}R_{\mu\nu\alpha\beta}\gamma^{\alpha\beta}\right)\left(-\bar{\psi}_\rho\g^{\mu\nu\rho}\chi+\bar{\chi}\g^{\mu\nu\rho}\psi_\rho +i\bar{\chi}\gamma^{\mu\nu}\chi\right)\\
  \mathcal{L}_{D^3} &= \frac{i}{m^2}\bar{\chi}\gamma^{\mu\nu\rho} D_\mu D_\nu D_\rho\chi=-\frac{1}{2m^2}\bar{\chi}\gamma^{\mu\nu\rho} 
  \left( qF_{\mu\nu}-\frac{i}{4}R_{\mu\nu\alpha\beta}\gamma^{\alpha\beta}\right)
  D_\rho\chi\notag\\ 
  &= i\bar{\chi}\gamma_\mu
  \left( \frac{1}{2m^2}G^{\mu\nu}-\frac{q}{m^2}\tilde{F}^{\mu\nu} \gamma^5 \right)
  D_\nu\chi
  .\label{LD3}
\end{align}
In the St\"uckelberg formulation, the problematic algebraic constraint is replaced by a gauge redundancy, so the relevant characteristic analysis is captured most transparently by the $\chi$ sector.
Equations of motion for $\chi$ are
\begin{align}
    \frac{3}{2}i\sda\chi+\frac{q}{m^2}\gamma_\mu\tilde{F}^{\mu\nu}\gamma^5\partial_\nu\chi+
   \frac{i}{2m^2}\gamma_\mu G^{\mu\nu}\partial_\nu \chi +\hdots =0,
\end{align}
where we included only the principal part.
The equation for $n_\mu$, the normal to the characteristic hypersurface, is obtained by setting the following determinant to zero:
\begin{align}
   \mathcal{M} 
   &=\gamma_\mu \left(n^\mu +\frac{1}{3m^2}G^{\mu\nu} n_\nu+\frac{2q}{3m^2}\tilde{F}^{\mu\nu} \gamma^5n_\nu\right).
\end{align}
The determinant vanishes when
\begin{align}
    \left( n^\mu +\frac{1}{3m^2}G^{\mu\nu} n_\nu\pm \frac{2q}{3m^2}\tilde{F}^{\mu\nu} n_\nu\right)^2=0 \; ,
\end{align}
which is the same equation as \eqref{determinant_gravity_EM}.

\section{Propagator and gauge fixing}\label{sec:propagator}
\subsection{Unitary propagator at high momentum}
In our discussion, we have considered the massive Rarita-Schwinger field as a gauge system together with an additional St\"uckelberg field $\chi$. This form is required to obtain the propagator that falls with momentum as $1/p$, which is needed for standard perturbative power counting.

We start with the free massive RS field in flat spacetime whose Lagrangian density is
\begin{align}
\mathcal{L}_{}=-i\Bar{\psi}_\mu \g^{\mu\nu\rho}\partial_\nu \psi_\rho+m\bar{\psi}_\mu\gamma^{\mu\nu}\psi_\nu ,\end{align}
in the momentum space this is
\begin{align}
\mathcal{L}_{}=\Bar{\psi}_\mu \g^{\mu\nu\rho}p_\nu \psi_\rho+m\bar{\psi}_\mu\gamma^{\mu\nu}\psi_\nu.
\end{align}
The inverse of the kinetic operator is the propagator \cite{Freedman:1976aw}\footnote{Obviously, this propagator has no limit for $m\rightarrow 0$, but the well known 
Van Dam-Veltman discontinuity for $m\rightarrow 0$ can be recovered by sandwiching it 
between two conserved spinor currents \cite{Deser:2000de, Duff:2002sm}. By contrast,
the renormalizable propagators (71) do admit a smooth limit for $m\rightarrow 0$,
so the discontinuity simply boils down to the fact that a massive gravitino
retains four helicities in the massless limit, instead of only two as appropriate for a
massless gravitino.} 
\begin{align}\label{unitary_propagator}
   \left[ \g^{\nu\rho\mu}p_\rho+m\gamma^{\nu\mu}\right] ^{-1}&=\frac{\slashed{p} +m}{p^2-m^2}\left[\eta^{\mu\nu}-\frac{1}{3}\g^\mu\g^\nu+\frac{p^\mu\g^\nu}{  3m}-\frac{\g^\mu p^\nu}{  3m}
    -\frac{2p^\mu p^\nu}{3m^2}\right].
\end{align}
This form is not appropriate for a quantum theory as it contains positive powers of momentum, which worsens the superficial degree of divergence and obscures standard power counting in loop calculations.

In the absence of the St\"uckelberg spinor $\chi$, i.e. when it is set to zero, this action takes the usual unitary gauge form whose propagator contains positive powers of momentum. To obtain a kinetic term with propagator $\sim 1/p$, we have to add a gauge-fixing term and include ghosts to preserve a proper number of propagating degrees of freedom.

\subsection{Gauge fixing}

Let us consider a gauge fixing condition that leads to a simple gauge-fixed Lagrangian:
\begin{align}
    G&=\g\cdot\psi +i\chi, \qquad \frac{\delta G}{\delta \epsilon}=i(-i
    \sda-m),\\
    \bar{G}&= \bar{\psi}\cdot\gamma-i\bar{\chi}, \qquad \frac{\delta \bar{G}}{\delta \bar{\epsilon}}=-i(-i
    \sda-m),\\
    \mathcal{L}_{\mathrm{gf}} &= \frac{3}{4}\left(\bar{\psi}\cdot\gamma-i\bar{\chi}\right)(i\sda -m)\left(\g\cdot\psi +i\chi\right) \label{eq:gauge_fixing},
\end{align}
where gauge fixing functions $G,\bar{G}$ are chosen to generalize $\gamma\cdot \psi$ gauge condition from the massless case.
This gauge choice was used for a massive gravitino in AdS in \cite{Duff:2002sm} and is analogous to the gauge $\frac{1}{4}\bar{\psi}\cdot\g\sda\g\cdot \psi$ from massless supergravity \cite{Das:1976ct, Sterman:1977ds}.
Then the gauge fixed Lagrangian takes the form
\begin{align}
    \mathcal{L}+\mathcal{L}_{\mathrm{gf}} &=\mathcal{L}_{\psi}
    +\frac{9}{4}\bar{\chi}(i\sda +m)\chi+\frac{3}{4}im(\bar{\psi}\cdot \g\chi-\bar{\chi}\g\cdot \psi)-\frac{3}{4}(\bar{\psi}\cdot \g\sda\chi-\bar{\chi}\sda\g\cdot \psi)+ \frac{3}{4}\bar{\psi}\cdot\gamma(i\sda -m)\g\cdot\psi \\
    &=\mathcal{L}_{\psi}
    +\left(\frac{1}{2}\bar{\psi}\cdot\g-\frac{3}{2}i\bar{\chi}\right)(i\sda +m)\left(\frac{1}{2}\g\cdot \psi+\frac{3}{2}i\chi\right)+ \frac{1}{2}\bar{\psi}\cdot\gamma(i\sda -2m)\g\cdot\psi \label{gauge_fixed_lagr_flat}.
\end{align}

The Lagrangian \eqref{gauge_fixed_lagr_flat} can be put into a diagonal form with Dirac kinetic terms. Consider the field redefinition \cite{Endo:1985km}
\begin{align}
    \psi_\mu = \phi_\mu -\frac{1}{2}\g_\mu \gamma\cdot \phi,
\end{align}
which flips the sign of the $\gamma$-trace component 
\begin{align}
    \gamma\cdot \phi =-\gamma\cdot \psi.
\end{align}
Then
\begin{align}
   -i\Bar{\psi}_\mu \g^{\mu\nu\rho}\partial_\nu \psi_\rho+m\bar{\psi}_\mu\gamma^{\mu\nu}\psi_\nu=-\bar{\phi}_\mu(i\sda+m)\phi^\mu-\frac{1}{2}\bar{ \phi}\cdot \gamma(i\sda-2m)\gamma\cdot \phi,
\end{align}
and the gauge-fixed Lagrangian takes a simple form
\begin{align}\label{L_gf_phi}
 \mathcal{L}+\mathcal{L}_{\mathrm{gf}} &=-\bar{\phi}_\mu(i\sda+m)\phi^\mu 
    +\bar{\xi} (i\sda +m)\xi 
    ,
\end{align}
where
\begin{align}
    \xi= \frac{1}{2}\g\cdot \psi+\frac{3}{2}i\chi=-\frac{1}{2}\g\cdot \phi+\frac{3}{2}i\chi.
\end{align}
The Lagrangian \eqref{L_gf_phi} leads to a very simple propagator structure for perturbation theory as
\begin{align}\label{fall_prop}
    \langle \phi_\mu \bar{\phi}_\nu\rangle=  -\eta_{\mu\nu}\frac{i\spa-m}{p^2-m^2+i0},\qquad  \langle \xi \bar{\xi}\rangle = \frac{i\spa-m}{p^2-m^2+i0}.
\end{align}
 However, we have to compensate the gauge freedom by introducing ghosts. 

\subsection{Ghost action}
In this section, we derive the ghost action in an explicitly BRST-invariant way.
We introduce a complex ghost spinor $c$ and its Dirac adjoint $\bar{c}$, associated with fermionic gauge parameter $\epsilon, $ and its adjoint $\bar{\epsilon}$, so BRST transformations on the minimal set of fields $(\psi_\mu,\bar{\psi}_\mu,\chi,\bar{\chi},c,\bar{c} )$ take the form
\begin{align}
    s\psi_\mu = \left( \partial_\mu-i\frac{m}{2}\gamma_\mu\right)c,\  s\bar{\psi}_\mu =\bar{c} \left( \overleftarrow{\partial}_\mu+i\frac{m}{2}\gamma_\mu\right),\
   s \chi =mc,  \ s \bar{\chi} =m\bar{c},\ sc=0,\ s\bar{c}=0,\end{align}
and gauge-fixing functions transform as
\begin{align}
       sG =\left(\sda -im\right)c,\qquad  s\bar{G} =-\bar{c}\left(\overleftarrow{\sda} +im\right).
\end{align}
Here $c,\bar{c}$ are commuting spinor ghosts.
To the set of fields we add a complex antighost $c', \bar{c}'$ and Nakanishi–Lautrup auxiliary anticommuting spinor fields $b,\bar{b}$, which transform 
\begin{align}
     sc'=b,\quad s\bar{c}'=\bar{b}, \quad sb=0,\quad  s\bar{b}=0.
\end{align}
The appearance of a dynamical Nakanishi–Lautrup field  $b$ is related to the fact that there is a kinetic
 term $(i\sda -m)$ between gauge fixing functions in \eqref{eq:gauge_fixing}. This additional ghost is required to obtain the proper number of degrees of freedom\footnote{The fields $\psi_\mu, \chi, b'$ contribute twelve spinor polarizations before the gauge redundancy is divided out. The two commuting spinor ghosts $c,c'$, whose fourth-order kinetic operator is equivalent to two Dirac-type spinor determinants, remove eight of these degrees of freedom. The net result is therefore equivalent to four physical polarizations, as appropriate for massive spin-$\frac{3}{2}$.}
and is related to Nielsen-Kallosh ghost \cite{Nielsen:1978mp, Kallosh:1978de}.

We aim to write a BRST-closed total action with a Gaussian gauge-fixing term \eqref{eq:gauge_fixing}.
For this reason we introduce the gauge-fixing fermion 
\begin{align}
    \Xi &=\frac{3}{4}\int d^4x \left[\bar{c'} (i\sda -m)\left(G -\frac{1}{2}b\right)+\left(\bar{G} -\frac{1}{2}\bar{b}\right) (i\sda -m)c' \right ],
    \end{align}
   which describes both gauge-fixing terms and ghost action and the sum is BRST-exact
\begin{align}
    S_{\mathrm{gf+gh}} = s\Xi, 
\end{align}
which is equal to
\begin{align}
    s\Xi =&\frac{3}{4}\int d^4x  \Bigg[\bar{b} (i\sda -m)\left(G -\frac{1}{2}b\right)+\left(\bar{G} -\frac{1}{2}\bar{b}\right) (i\sda -m)b \notag\\&+ \bar{c'} (i\sda -m)\left(\sda -im\right)c- \bar{c}(\sda -im) (i\sda -m)c' \Bigg]\\
  =&\frac{3}{4}\int d^4x  \Bigg[\bar{G} (i\sda -m)G-\left(\bar{G} -\bar{b}\right) (i\sda -m)\left(G -b\right) \notag\\ &+ \bar{c'} (i\sda -m)\left(\sda -im\right)c- \bar{c}(\sda -im) (i\sda -m)c' \Bigg].
\end{align}
The term $\left(G -b\right)$ can be either shifted or integrated to a functional determinant and reintroduced with a spin-$\frac{1}{2}$ field $b'$,
and ghost $c,c'$ can be rescaled, so the total action is
\begin{align}\label{ghosts_free_action}
    S_{\mathrm{tot}} =& S_{\mathrm{inv}} +s\Xi
    =\int d^4x  \Big\{-\bar{\phi}_\mu(i\sda+m)\phi^\mu
    +\bar{\xi} (i\sda +m)\xi  \notag\\ &-\bar{b}' (i\sda -m)b'+ \bar{c'} (i\sda -m)\left(\sda -im\right)c- \bar{c}(\sda -im) (i\sda -m)c' \Big\}.
\end{align}
This is the complete gauge-fixed action in the free theory.
Minimal interactions with gravity and gauge fields are introduced by covariantizing derivatives at each step and lead to the following BRST-invariant action 
\begin{align}\label{eq:total_action}
    S_{\mathrm{tot}} 
    =&\int d^4x  \sqrt{-g} \Big\{-\bar{\phi}_\mu(i\sDa+m)\phi^\mu 
    +\bar{\xi} (i\sDa +m)\xi+\mathcal{L}_{D^2}+\mathcal{L}_{D^3} \notag \\
    &-\bar{b}' (i\sDa -m)b' + \bar{c'} (i\sDa -m)\left(\sDa -im\right)c- \bar{c}(\sDa -im) (i\sDa -m)c'\Big\},
\end{align}
where $\mathcal{L}_{D^2}+\mathcal{L}_{D^3}$ are higher order curvature terms \eqref{LD2}-\eqref{LD3}, and $D$ acts on spinor components of $b,c,c'$ as in \eqref{covariant_derivatives_CST}.  
The total BRST-invariant action \eqref{eq:total_action} provides the starting point for perturbative calculations around flat spacetime. The quadratic part has the Dirac-like form \eqref{L_gf_phi}, ensuring that the free propagators \eqref{fall_prop} fall as $1/p$ at large momenta. The interactions with the gravitational and electromagnetic backgrounds are introduced by covariantizing the derivatives, producing the higher-order curvature couplings  $\mathcal{L}_{D^2}$, $\mathcal{L}_{D^3}$, which are suppressed by $m^{-1}$ and $m^{-2}$. For gravitinos near the Planck scale, these corrections are suppressed by inverse powers of $M_{\mathrm{ Pl}}$. The commuting spinor ghosts $c, c'$ carry the same charge and spin quantum numbers as the gauge parameter $\epsilon$ and, together with the anticommuting Nakanishi–Lautrup field $b'$, restore the net degree-of-freedom count off-shell.

\section{Conclusions}

We have revisited the consistency conditions for a charged massive Rarita-Schwinger field in Einstein-Maxwell backgrounds from the perspective of a superheavy charged gravitino. The classical constraint and characteristic analyses lead to the same local condition: in the minimally coupled theory, any Einstein-Maxwell background is free of the Velo-Zwanziger obstruction provided the inequality \eqref{inequality} holds.
This bound is due to Deser and Waldron \cite{Deser:2001dt} but our interpretation differs: for ordinary low-energy phenomenology the Planckian scale of the bound is an obstruction, whereas for fractionally charged gravitino dark matter it is precisely the regime of interest.

One may ask whether the bound remains valid for a general effective theory of massive charged spin-$\frac{3}{2}$, in particular when the coupling is nonminimal. This changes equations of motion and hence also equations for characteristic surfaces. For hermitian nonminimal Pauli-like interactions of the form
\begin{align}
\mathcal{L}_{\mathrm{Non-min}} =    \frac{iq}{m}\bar{\psi}_\mu\left( aF^{\mu\nu} +ib\gamma_5\tilde{F}^{\mu\nu}
\right)\psi_\nu,
\end{align}
the bound  \eqref{inequality} is modified by factors of order $\mathcal{O}(a,b)$, 
but the conclusion remains the same: causal propagation can be restored for Planck scale masses (see \cite{Deser:2000dzED, Deser:2001dt} for discussion of expressions obtained this way).
Nonminimal couplings are crucial in supergravity, and it is important that their presence preserves, up to a numerical coefficient, our conclusion about the Planckian mass scale.
Interestingly, the  BPS and causality bounds coincide for the theory with supergravity nonminimal coupling with Pauli term
\begin{align}
\mathcal{L}_{\mathrm{SUGRA}} =    \frac{iq}{m}\bar{\psi}_\mu\left( F^{\mu\nu} -i\gamma_5\tilde{F}^{\mu\nu}
\right)\psi_\nu.
\end{align}
Deser and Waldron found that then the causality bound is \cite{Deser:2001dt} 
\begin{align}
 m^2> M_{\mathrm{SUGRA}}^2=8\pi\alpha_e \frac{q^2}{e^2} M_{\mathrm{Pl}}^2   \; ,
\end{align}
which is equal to the BPS bound \eqref{BPS_bound}.

We also reformulated the system using a St\"uckelberg spinor. In this language, the algebraic constraint of the massive Rarita-Schwinger field is traded for a fermionic gauge redundancy, and the dangerous helicity-$\frac{1}{2}$ sector becomes explicit. The characteristic matrix of the St\"uckelberg spinor reproduces the same condition as the constrained Rarita-Schwinger analysis.

Finally, we discussed the propagator. The usual massive Rarita-Schwinger propagator has unitary-gauge high-momentum behavior and is not appropriate for standard perturbative power counting. The St\"uckelberg formulation allows a Feynman gauge in which the quadratic action is diagonal, has Dirac kinetic terms and leads to a renormalizable propagator. 

\vspace{1.5em}
\noindent{\bf Acknowledgments:} B.S.'s work was supported by a short term scientific mission grant from the COST action CA22113 THEORY-CHALLENGES. We also thank Karim Benakli for correspondence.

\printbibliography
\end{document}